\documentclass[conference]{IEEEtran}
\usepackage{blindtext, graphicx}
\usepackage[]{algorithm2e}
\usepackage[justification=centering]{caption}

\usepackage{multirow}
\usepackage[table]{xcolor}
\usepackage[hyphens]{url}
\usepackage[hidelinks]{hyperref}
\hypersetup{breaklinks=true}
\usepackage{cite}
\usepackage{afterpage}
\usepackage{lipsum}
\usepackage{caption}
\renewcommand{\baselinestretch}{1.125}

\ifCLASSINFOpdf
\else
\fi

\usepackage{geometry}
\begin{document}
%
\title{\huge Parallelizing Bisection Root-Finding:\\ A Case for Accelerating Serial Algorithms\\ in Multicore Substrates}

\author{\IEEEauthorblockN{Mohammad Bakhshalipour and Hamid Sarbazi-Azad}
\IEEEauthorblockA{\\Department of Computer Engineering, Sharif University of Technology\\School of Computer Science, Institute for Research in Fundamental Sciences (IPM)\\ \\
}}

\maketitle

\begin{abstract}
Multicore architectures dominate today's processor market. Even though the number of cores and threads are pretty
high and continues to grow, inherently serial
algorithms do not benefit from the abundance of cores and
threads. In this paper, we propose \textit{Runahead Computing}, a
technique which uses idle threads in a multi-threaded architecture
for accelerating the execution time of serial algorithms. Through
detailed evaluations targeting both CPU and GPU platforms and
a specific serial algorithm, our approach reduces the
execution latency up to $9x$ in our experiments.
\end{abstract}
\begin{IEEEkeywords}
Multi-Thread Programming, Single-Thread Performance, Multicore Processor, GPU, Bisection Root-Finding.
\end{IEEEkeywords}

%
\IEEEpeerreviewmaketitle

\section{Introduction}
Multi-threaded architectures presently appear across the whole spectrum of computing machines, from the low-end embedded processors to high-end general-purpose devices. Chip Multiprocessor (CMP)~\cite{olukotun1996case} is a type of multi-threaded processors, in which, cores do not share the computational resources. CMPs are implemented in many commercial systems and have high usage in broad classes of computations. Intel Nehalem i7~\cite{dixon2010next}, AMD Bulldozer~\cite{butler2011bulldozer}, IBM Power5~\cite{kalla2004ibm}, Sun Niagara T2~\cite{shah2007ultrasparc}, and TILE64~\cite{bell2008tile64} are examples of commercial CMPs. CMP enhances the performance of a single program when the program can be split into multiple pieces, each piece run by one core, in parallel with others. 

In the other side, Graphics Processing Units (GPUs) are also being adjusted for general-purpose computations. GPUs use an aggregate form of single-instruction-multiple-data (SIMD) paradigm~\cite{flynn1972some} with fine-grain multi-threading. NVIDIA Kepler GK110~\cite{nvidia2012nvidias} and AMD TeraScale~\cite{houston2008anatomy} are examples of commercial GPUs. GPU exploits the data-level parallelism of a single program which consists of the same operation on multiple data points for accelerating the execution time of the application~\cite{Sadrosadati:2018:LEH:3173162.3173211, Yazdanbakhsh:2015:NAG:2830772.2830810, regmutex, sadrosadati2017effective, aghilinasab2016reducing, karami2013statistical, abbasi2012preliminary, mirsoleimani2013parallel}. 

Even though the trend is growing the number of cores in both multicore and manycore systems~\cite{lotfi2012scale, ferdman2011cuckoo, grot2012optimizing, lotfi2010turbotag}, serial programs do not benefit from increasing the core count. Serial programs (or serial parts of a program) run on a single thread and cannot be split into two or multiple threads. Executing such programs on a manycore platform results in only one core running the whole program and other cores remaining idle. The situation gets worse when boosting the core count diminishes the single-core performance. The problem arises from two reasons: (1) Tight physical constraints of the chip (i.e., limited power and area budgets) prevent from accommodating hundreds of large and power-hungry cores on a single chip~\cite{esmaeilzadeh2011dark, hardavellas2011toward}. So increasing the core count implies replacing high-performance cores with simple and small cores, which leads to diminishing the performance of single-thread and consequently protracting execution time of serial programs. Some proposals resolve this problem with asymmetric architectures~\cite{suleman2009accelerating}. (2) Increasing the core count forces to replace non-scalable crossbars with on-chip networks which use scalable topologies (e.g., Mesh). Latency incurred by on-chip interconnect decreases the per-core performance due to slower accesses to shared caches. Boosting the core count grows the network hop count and results in longer delays~\cite{bakhshalipour2018fast, Mirhosseini:2017:BBN:3130218.3130222, Mirhosseini:2015:EVC:2757012.2757164, sadrosadati2015efficient, Lotfi-Kamran:2012:NMS:2457472.2457496, lotfi2017near, lotfi2010edxy, lotfi2008barp, lotfi2016efficient, lotfi2017noc, falsafi2017network, modarressi2011application, modarressi2010virtual, modarressi2009hybrid, modarressi2007power, modarressi2010efficient, mirhosseini2016quantifying, mirhosseini2017poster}. Some proposals offset this obstacle with richly-connected topologies~\cite{grot2011kilo, kim2007flattened} which have significant area and energy overheads.

In this paper, we propose \textit{Runahead Computing}, a technique for accelerating inherently serial algorithms on multicore and manycore platforms. Runahead Computing draws on previous researches on non-traditional parallelism and targets the execution latency of serial algorithms. The rest of the paper is organized as follows: Section {2} explains Runahead Computing and gives the essential background on non-traditional parallelism. In section {3}, we choose a proper serial algorithm as a case study and describe its details. Throughout section {4}, we illustrate the implementation of Runahead Computing in our case study. Section {5} discusses our evaluation methodology. Section {6} presents the results of evaluation experiments. Finally, section {7} concludes the paper.

\section{Runahead Computing}
Runahead Computing refers to use the idle threads in a multi-threaded architecture to improve the performance of the single-threaded application. Runahead Computing is a kind of speculation in which, idle threads operate few steps ahead of the main thread. These speculative threads attempt to provide a situation at their activity location, under which, when the main thread reaches there, it executes faster than usual.

This is not the first research on using idle thread contexts with the purpose of increasing the performance of the single-threaded application. Various schemes have been proposed to use idle thread contexts to provide certain kinds of assistance to the main thread\footnote{\hspace{1mm}This paradigm sometimes has been called non-traditional parallelism.}. Some approaches leverage from idle thread contexts as a prefetcher for main thread~\cite{collins2001dynamic, collins2001speculative, kim2002design, luk2001tolerating, zhang2005event, zilles2001execution} in order to defeat long-latency memory accesses~\cite{bakhshalipour2018domino, bakhshalipour2017efficient, vakil2018cache}. Several proposals precompute branch outcomes through a derived variant of main thread~\cite{zilles2001execution, chappell1999simultaneous}. There are also some approaches which exploit idle thread contexts to precompute dependent live-in data~\cite{madriles2008mitosis}. However, all of these approaches require significant non-trivial changes to the hardware of processors which makes their implementation challenging for shipping products. In this paper, we target a method that pushes everything to the software and hence is entirely applicable in the context of commercial off-the-shelf processors.

\begin{figure}[t] 
\includegraphics[width=0.4\textwidth]{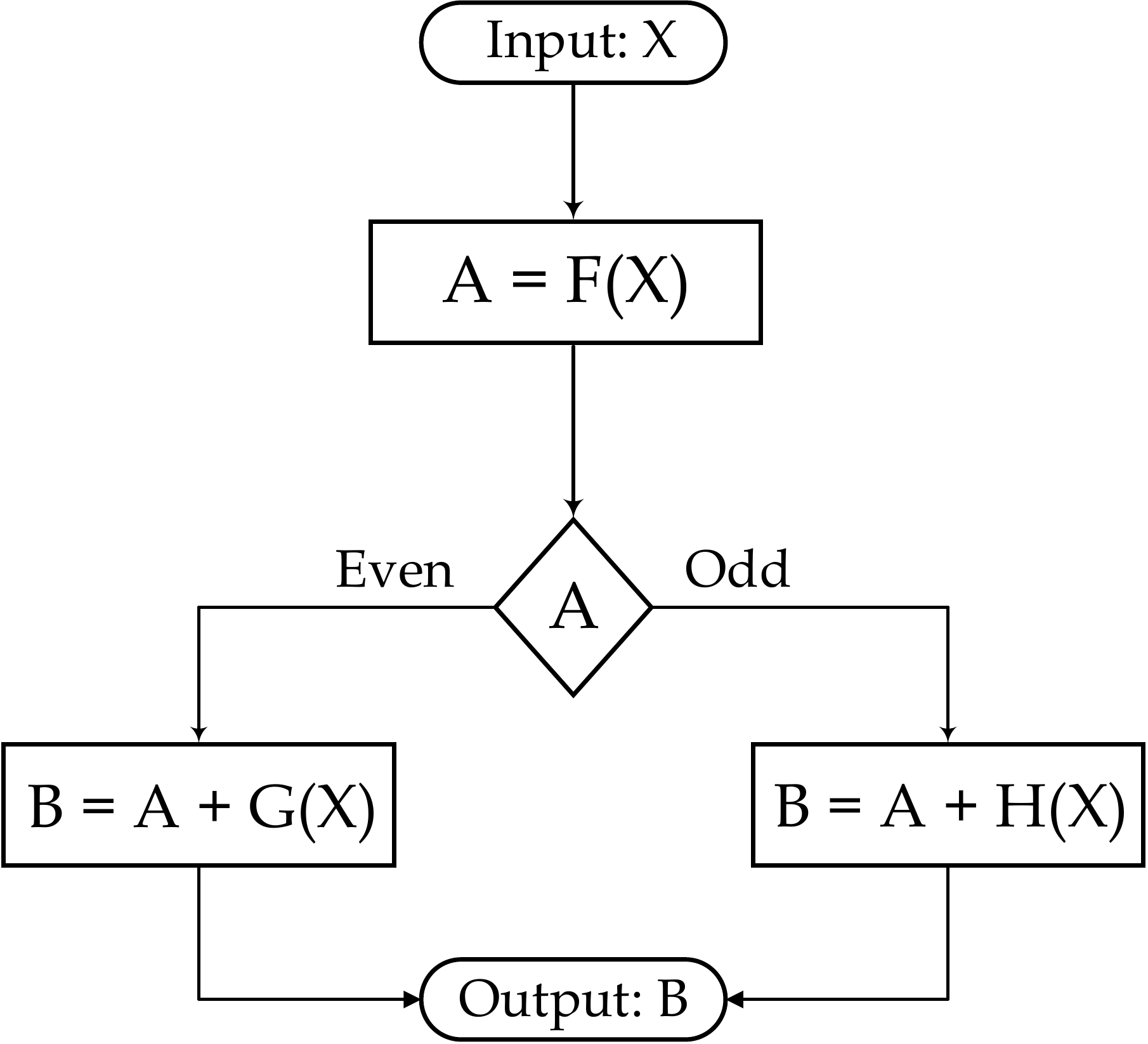}
\centering
\caption{Flowchart of a hypothetical program.}
\label{fig1:hypo}
\end{figure}

We proceed to describe our approach with an example. Figure~\ref{fig1:hypo} shows the flowchart of a hypothetical program. The program takes $X$ as input and returns $B$ as output. In the beginning, the program computes $F(X)$ and stores the result in variable $A$. Then, if $A$ is an even number, program proceeds with calculating $G(X)$ and storing the sum of the result and $A$ in variable $B$. But if $A$ is an odd number, the program computes the value of $H(X)$ and put the sum of the result and $A$ into variable $B$. Finally, it returns $B$ as the output of the program. Because of existing dependencies between parts of the program, it cannot be regularly parallelized. So considering the execution time of 10, 5, and 5 seconds respectively for $F$, $G$, and $H$, and neglecting the latency of other operations (e.g., store and add), a single thread can execute this program within 15 seconds. However, in the Runahead Computing, programmer initiates two threads at the beginning of execution. First thread (main thread) begins with calculating $F(X)$ and in parallel, the second thread (helper thread) calculates $G(X)$ and $H(X)$ and stores the result of them in variables $G$ and $H$, respectively. In this manner, the execution latency of calculating $G(X)$ and $H(X)$ overlaps with execution latency of computing $F(X)$. So, when the main thread finishes the calculation of $F(X)$, the results of $G(X)$ and $H(X)$ are ready in variables $G$ and $H$. So, the main thread checks the value of $A$ and picks one of $G$ or $H$ for summing with $A$ and storing the result in $B$ and returning the output. By this manner, the whole program finishes within 10 seconds. So, in this example, Runahead Computing improves the performance of the program by 50\%.\newline

%

\section{Bisection Method}

In this section, we choose \textit{Bisection} root-finding algorithm~\cite{corliss1977root} as our case study and describe its baseline manner (without Runahead Computing). Bisection is a serial algorithm and is a suitable case for our proposal due to its operative behavior.  

\subsection{Algorithm}
Bisection is a root-finding algorithm which operates on continuous functions. The bisection method is based on this lemma: If continuous function $f$ returns opposite sign values on $a$ and $b$, then equation $f(x) = 0$ has at least one real root in the interval $(a, b)$, where $a < b$. Bisection algorithm takes a function and an interval as inputs and repeatedly halves the interval and then picks a subinterval in which a root must lie.

\begin{figure}[t] 
\includegraphics[width=0.4\textwidth]{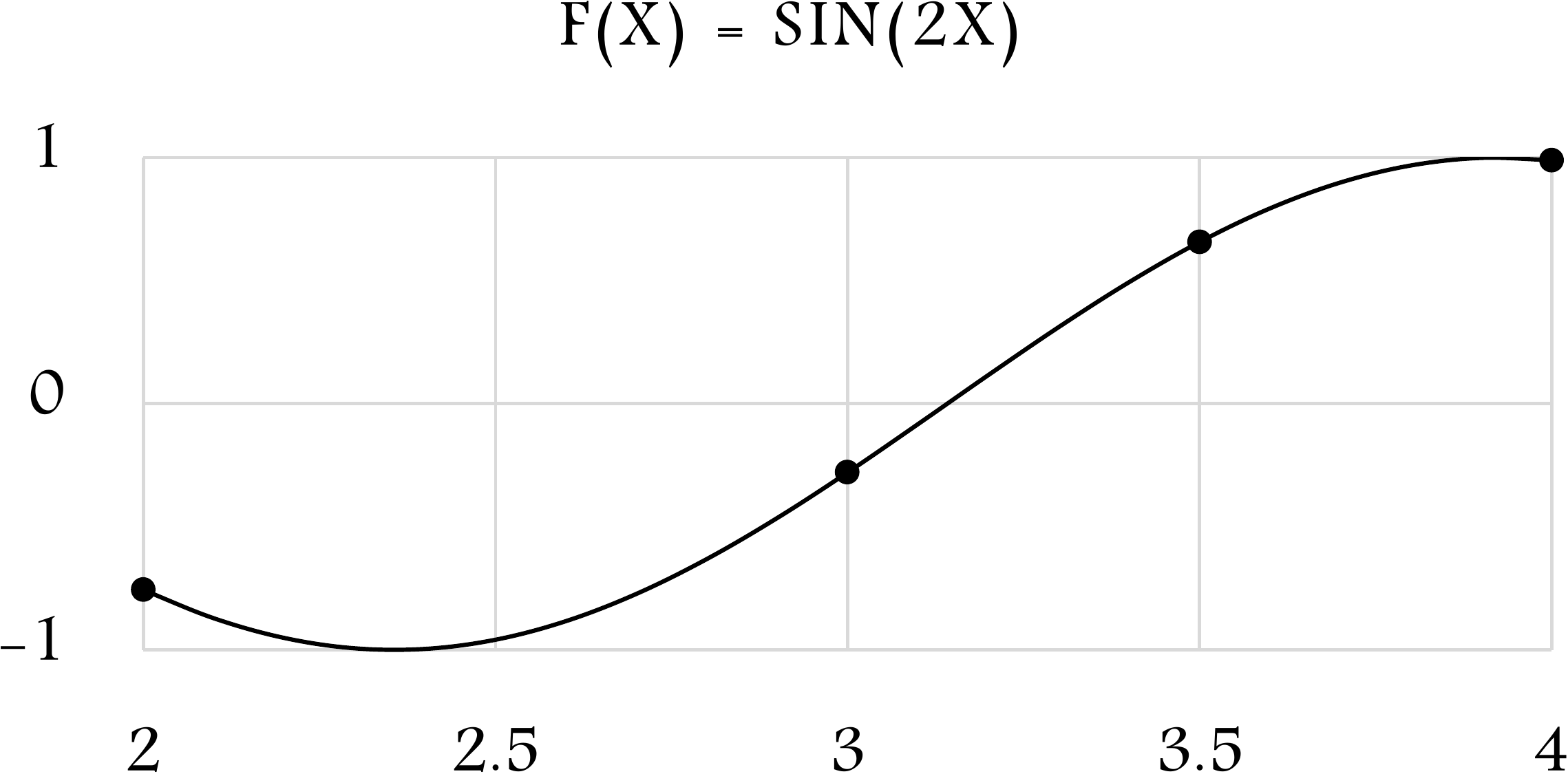}
\centering
\caption{Bisection algorithm on a sample function.}
\label{fig2:bisection}
\end{figure}

Figure~\ref{fig2:bisection} illustrates an example of the operation of this algorithm. The algorithm takes $f(x) = sin(2x)$ as input function and $(2, 4)$ as initial interval and then tries to find a root for equation $f(x) = 0$  in the given interval. At first, it computes $f(2)$ and $f(4)$ and finds that $f(2)$ is negative and $f(4)$ is positive. The algorithm computes the middle point of the interval and the value of the function at that point (i.e., $c = \frac{2+4}{2} = 3$ and $f(3)$) and finds that $f(3)$ is negative. Now, the interval should be halved and replaced by one of $(2, 3)$ or $(3, 4)$ intervals. Because the function returns opposite sign values on the edge points of $(3, 4)$ interval, the algorithm chooses $(3, 4)$ interval to continuation. In the next step, the operation repeats similarly and $(3, 3.5)$ interval is chosen. The algorithm continues in an alike manner for more iterations (not shown in the figure) till estimating the root with acceptable accuracy. Due to halving the interval at each step, in general, for achieving an error value less than $\epsilon$, we need to iterate $\lceil{\log_2 \frac{b-a}{\epsilon}}\rceil$ times for the initial interval of $(a, b)$.

The main advantage of Bisection algorithm is its simplicity and robustness. While other root-finding numerical methods, which may have higher performance, operate just when the function has specific conditions, Bisection method works on any continuous function regardless of any particular circumstances. The only problem with Bisection method is its high execution time which we target in this work. Some approaches have been proposed to use Bisection method for achieving a nearness to a solution which is formerly used as a starting point for more quickly converging methods~\cite{burden19852}. \newline

\subsection{Baseline Implementation}
We implement the Bisection method like Algorithm~\ref{alg:bisection-serial} for baseline evaluations. Throughout this implementation, even if the exact root is found in the middle of the execution, the program does not stop and continues for the predefined number of iterations.

\begin{algorithm}[b]
 \vspace{5mm}
 \KwData{Function: $f$, Interval: $(a, b)$, Iterator: $iterations$}
 \KwResult{root of $f(x) = 0$ in $(a, b)$}
 \vspace{1mm}
 \While{$iterations > 0$}{
  \vspace{2mm}
  $iterations \gets iterations -1$\\
  $root \gets \frac{a + b}{2}$\\
  \eIf{$f(a) \times f(root) < 0$}{
   $b \gets root$\\
   }{
   $a \gets root$\\
  }
 }
 \Return $root$\\ \vspace{5mm}
 \caption{Baseline implementation for Bisection algorithm.}
 \label{alg:bisection-serial}
\end{algorithm}

\section{Runahead Bisection}

In this section, we propose our approach for accelerating the Bisection algorithm by leveraging available idle threads. First, we describe our method in detail and then discuss its complexity.

\subsection{Algorithm}
We begin with considering the same example (i.e., Fig.~\ref{fig2:bisection}) in an environment with three threads. One of the threads is the main thread, and the two others are helper threads. At first, the main thread computes the value of $f(3)$, and in parallel, helper threads operate on one step ahead of the main thread. One helper thread predicts that $f(3)$ will be positive and begins to compute the value of $f(2.5)$, speculatively. The other helper thread predicts a negative value for $f(3)$ and speculatively calculates the amount of $f(3.5)$. Each helper thread stores the result of its computation in a shared variable (say $f_{3.5}$ and $f_{2.5}$). So, when the main thread completes its calculation, it compares the result of $f(3)$ with the results of helper threads which have been stored in two shared variables. Because $f(3)$ is negative (i.e., the prediction of second helper thread is correct) and $f_{3.5}$ is positive, the next interval is $(3, 3.5)$. In the next step, the main thread computes $f(3.25)$, and the helper threads calculate $f(3.125)$ and $f(3.375)$. Operations repeat the subsequent steps similarly. In this fashion, if we ignore the latency of some operations (e.g., joining threads and storing the values), the execution latency reduces by 50\%.

We can further reduce the execution latency of the algorithm with devoting more helper threads to the program. For example, if we have seven threads, we can do the computations related to the two steps ahead, in the current stage. In this situation for our example, initially, the main thread computes $f(3)$, two helper threads compute $f(3.5)$ and $f(2.5)$, and four helper threads compute $f(2.25)$, $f(2.75)$, $f(3.25)$, and $f(3.75)$. By this way, the execution latency drops to one-third of the baseline latency.

\begin{figure}[t] 
\includegraphics[width=0.4\textwidth]{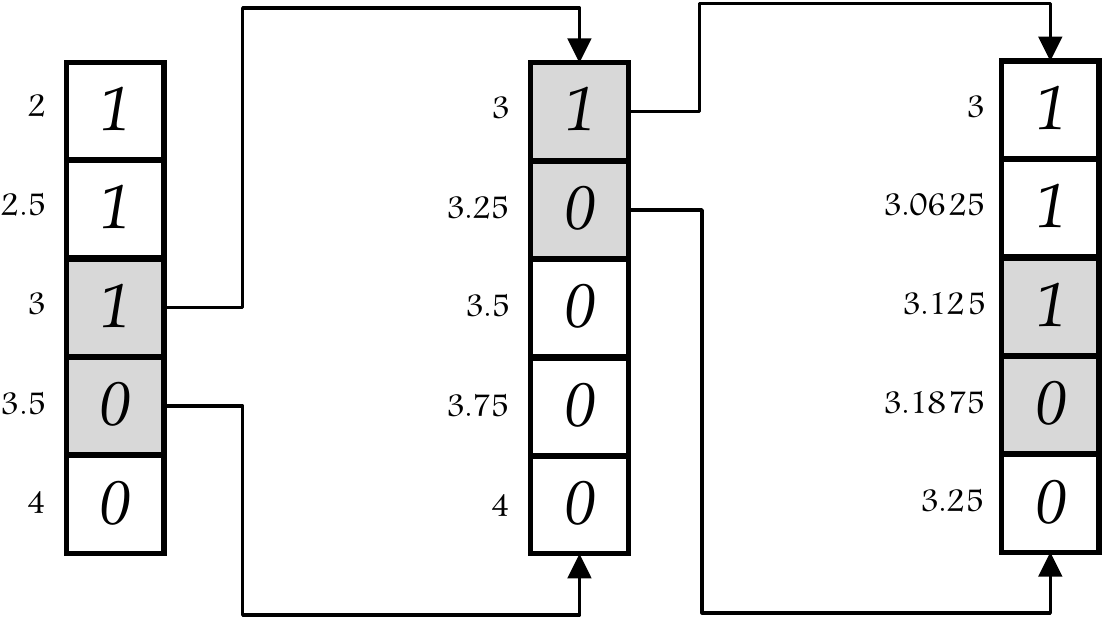}
\centering
\caption{Array-based implementation of Runahead Bisection.}
\label{fig3:array}
\end{figure}

For preserving the scalability of our method to the number of threads, we implement the shared variables as an array. Each thread has a particular element in the array. The array also contains two elements which do not belong to any thread and show the sign of edge points' value of the current interval. Any thread fills the corresponding element in the array after its computation finishes. If the result of the computation is negative, the thread fills the corresponding element in the array with `1'. Otherwise, it sets the element to `0'. Whenever all the threads write the results into the array, each thread compares the entries in the array which belong to the two neighbor threads (it is a simple XOR). If they are not the same, this means one edge of the new interval is the point of this thread, and the other edge is the neighbor point which has a different value in the array. Based on this, the new interval is chosen, then the main thread begins with computing the value of middle-point of the new interval, and helper threads pick their corresponding points, and the scenario repeats like before. To prevent false-sharing~\cite{torrellas1994false, jeremiassen1995reducing}, we implement the array as a two-dimensional structure, but we only use one dimension\footnote{\hspace{1mm}In this manner, shared variables map to different coherence units.}. Figure~\ref{fig3:array} illustrates the array-based implementation of Runahead Bisection for our example with three threads. At first, main thread computes $f(3)$ and writes `1' to the array, as the result is negative. Parallel with the main thread, helper thread-1 computes $f(2.5)$, and helper thread-2 calculates $f(3.5)$; then, they write their results (`0' or `1') into the array. Now, the array is complete\footnote{\hspace{1mm}Formerly, we know the signs of $f(2)$ and $f(4)$.}. Helper thread-1 compares sign of the value of its neighbor points (i.e., $f(2)$ and $f(3)$). In parallel, main thread and helper thread-2 do similar work on their corresponding entries in the array. Based on comparisons, helper thread-2 finds that its neighbors have different-sign values. So, for next step, helper thread-2 sets the interval (which is a shared variable among all threads) to $(3, 3.5)$. The sign of the value of edge points gets copied to the top and bottom of the array, and the other operations repeat in the previous fashion.\newline

\setcounter{footnote}{0}
\renewcommand{\thefootnote}{\alph{footnote}}

\subsection{Complexity}
Because of similarity between operations of all threads, we first analyze the time complexity of a single thread. Each thread computes the value of its corresponding point; then it writes the sign of result in the array; afterward, it waits for synchronizing with other threads. Next, the thread compares the sign of results of neighbor threads and updates the interval if required. All of these operations take $\mathcal{O}(1)$ latency (as the execution times of these operations are constant and independent of demanded accuracy or the length of the initial interval). 

If we need to iterate $n$ times to reach to a certain accuracy in the baseline algorithm, in the Runahead manner, we need to iterate $\frac{n}{k}$ times for each thread, where $k$ depends on the number of threads. Generally, we can reduce the number of iterations for each thread, from $n$ to $\frac{n}{k}$, if we have $2^{k}-1$ threads. So, given $k$ threads, our approach reduces the total execution time complexity of the program from $\mathcal{O}(n)$ to $\mathcal{O}(\frac{n}{log_{2}(k+1)})$, where $n$ is the number of required iterations in the baseline implementation.

\section{Methodology}
We evaluate our approach on both CMP and GPU. Table~1 summarizes the parameters of our platforms. For CPU, we compile the program using GCC without optimization. For GPU, we use NVCC for compilation, again without optimization. For eliminating the effects of compulsory cache misses, we run each program two times and report the results of the second execution. To demonstrate the effectiveness of our method, we choose the function as a high-latency function. We use trigonometric functions and calculate them with Taylor series~\cite{ahlfors1953complex}.

\vspace*{13pt}

\vspace*{-12pt}

\begin{table}[b!] 
 \begin{center} 
  \caption{Evaluation parameters.} 
  \label{table:method} 
    \begin{tabular}{| l | l |} 
     \hline 
       \rowcolor{gray!20}
       {\bf Parameter}                    & {\bf{Value}} \\ 
     \hline 
     \hline 
       {CPU}				  & x86 Architecuter, Intel Core i7, 2.4 GHz, Eight cores\\  

     \hline 
       OS               & Linux, Kernel version: 4.4.0-34 \\ 

     \hline 
       GPU     & NVidia Tesla K20, 732 MHz\\

      \hline 
       \multirow{2}{*}{Program}  	  & $f(x) = \sin(\cos(x))$, Taylor series with $10^{4}$ iterations\\  
					  & Initial interval: $(1, 2)$\\ 

     \hline 
    \end{tabular} 
 \end{center} 
\end{table}

\section{Evaluation Results}
In this section, we report the results of two sensitivity analysis: (1) Sensitivity of execution time to the number of threads, and (2) Sensitivity of speed-up of our method to the latency of input function. 

\subsection{Sensitivity to The Number of Threads}
For a specific input program which has been defined in Table~1, we sweep the number of threads and measure the execution time of the application. For CPU program, we set the maximum tolerable error to $2^{-6}$ and sweep the number of threads from $1$ to $7$. For GPU program, we set the maximum acceptable error to $2^{-2520}$ and sweep the number of threads from $1$ to $1021$. We choose $2^{-6}$ and $2^{-2520}$ as the maximum tolerable errors because, in these situations, the number of iterations of the single-threaded program is divisible by $\log_{2}(\#$Threads$+1)$.

\begin{figure}[t] 
\includegraphics[width=0.4\textwidth]{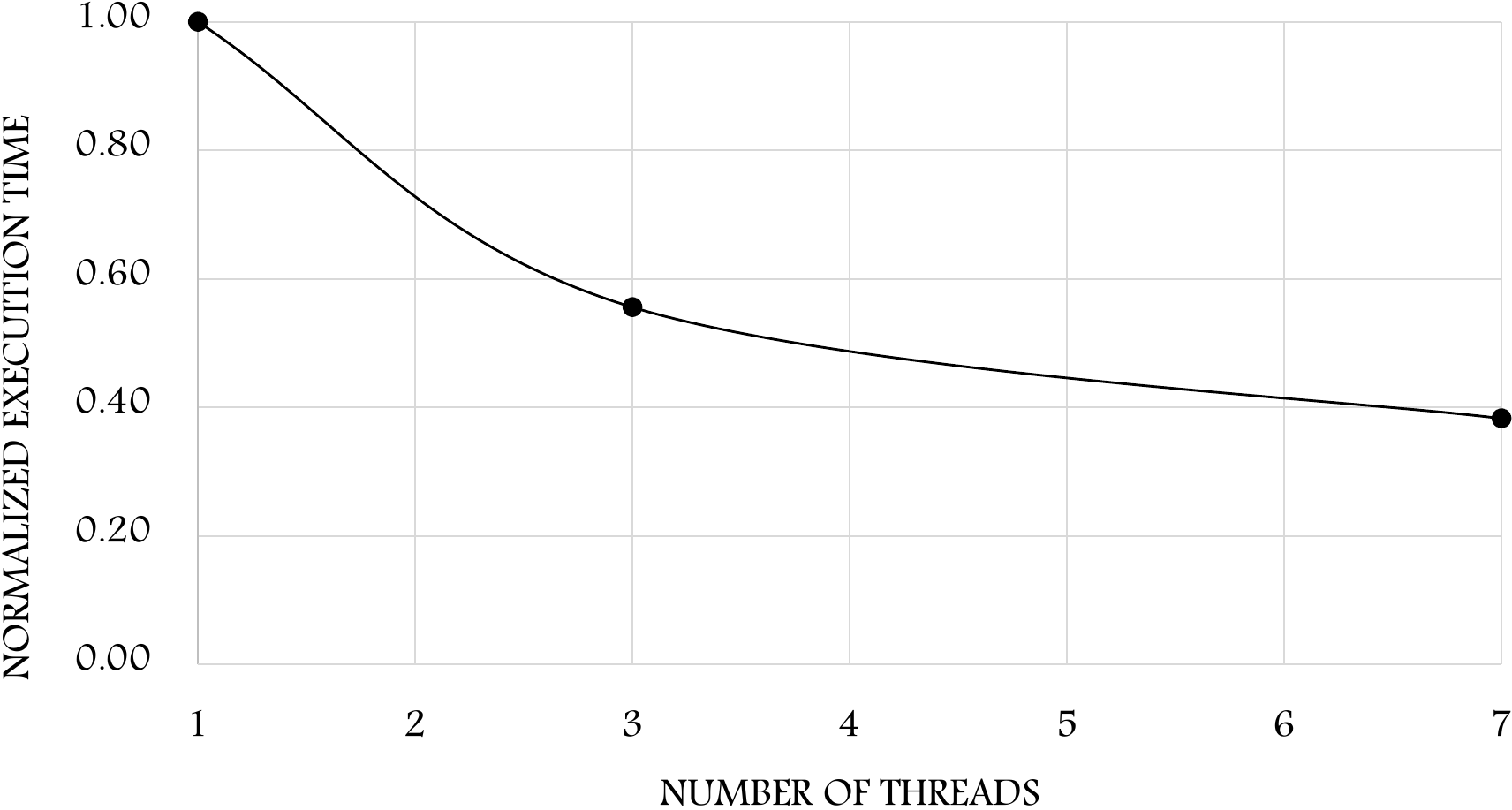}
\centering
\caption{Sensitivity of execution time of CPU program to the number of threads.}
\label{fig4:cpu-exec}
\end{figure}

Figure~\ref{fig4:cpu-exec} shows the result of thread-sweeping on CPU program. The execution latency values are normalized to that of the single-threaded program. The execution time of CPU program drops to 55\% (using three threads) and 38\% (using seven threads) of its baseline serial implementation. As the figure illustrates, the performance nearly scales with increasing the thread count. The noises in the scaling come from the latency of operations which we do just in the multi-threaded program (e.g., creating and synchronizing the threads, filling the variables which are shared among threads). Notably, by increasing the thread count, the latency of these operations (e.g., synchronizing the threads) increases and prevents reaching perfect performance scaling. \newline

\begin{figure}[b]
\includegraphics[width=0.4\textwidth]{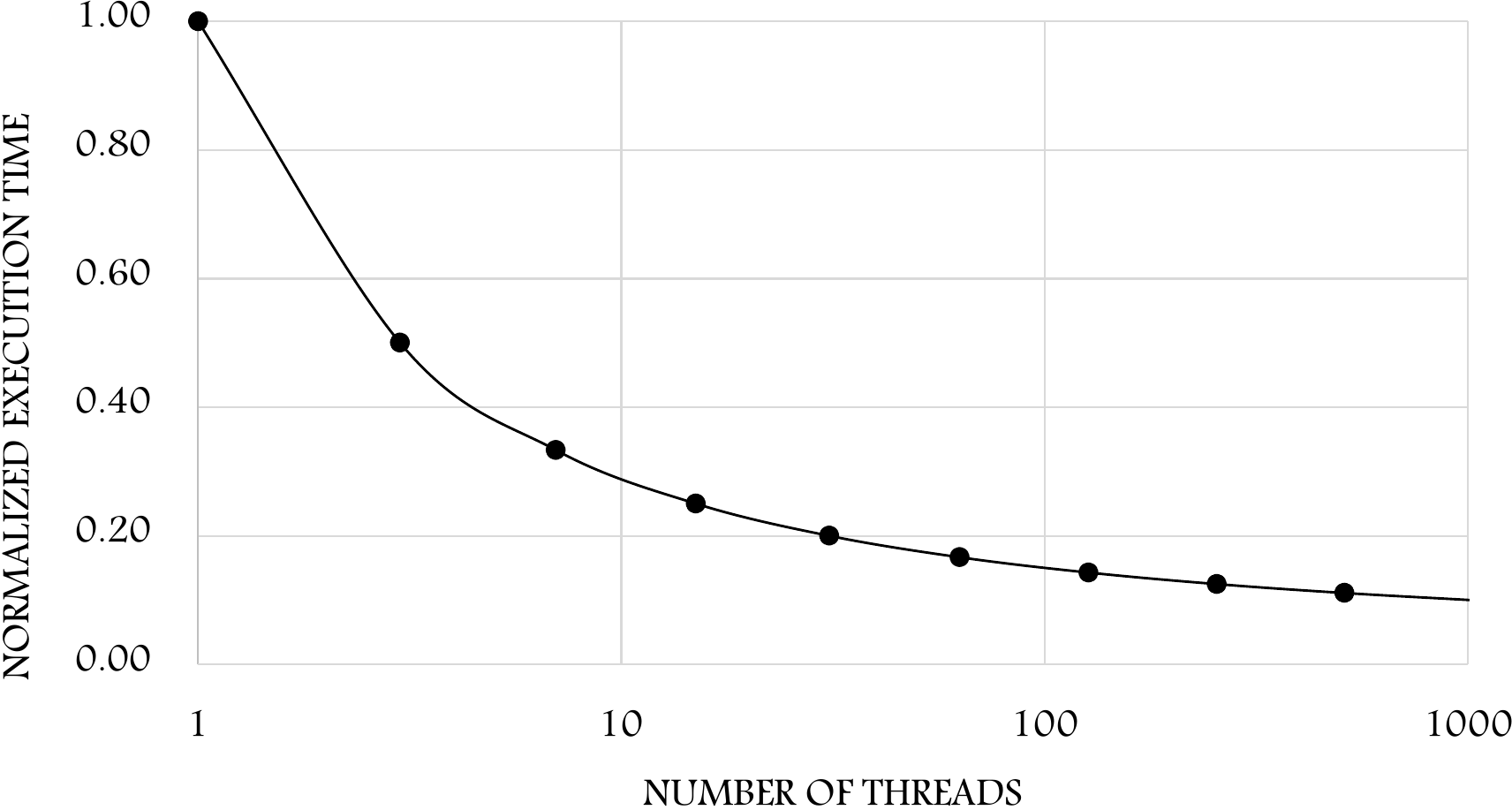}
\centering
\caption{Sensitivity of execution time of GPU program to the number of threads.}
\label{fig5:gpu-exec}
\end{figure}

Figure~\ref{fig5:gpu-exec} presents the result of sweeping the number of threads on the execution latency of the GPU program. Again, the execution times are normalized to that of the single-threaded program. The latency of program drops to 50\% (using three threads) and 10\% (using 1023 threads) of its baseline serial implementation. As the figure confirms, the performance perfectly scales with growing the number of threads. The low overhead of creating/joining hardware threads in the GPU platform~\cite{nvidia2008programming} provides this near-ideal scalability. \newline

\subsection{Sensitivity to The Execution Latency of Input Function}
In this section, we investigate the sensitivity of our approach's speed-up to the computation latency of input function. For this reason, we sweep the number of iterations of Taylor series for computing trigonometric functions. By growing the number of iterations in Taylor series, the latency of calculating $\sin(x)$ and $\cos(x)$ raises, and consequently, the entire latency for computing the value of a given point grows. In this experiment, we set the maximum tolerable error to $2^{-6}$ and restrict the number of threads to three. By this way, the single-threaded program needs to iterate six times, and the multi-threaded program requires three iterations. In an ideal situation (i.e., neglecting the latencies of creating and joining threads and filling shared variables), the multi-threaded application should take the half of execution time of the single-threaded program (in other words, multi-threading should raise the performance by 100\%). 

\begin{figure}[t]
\includegraphics[width=0.4\textwidth]{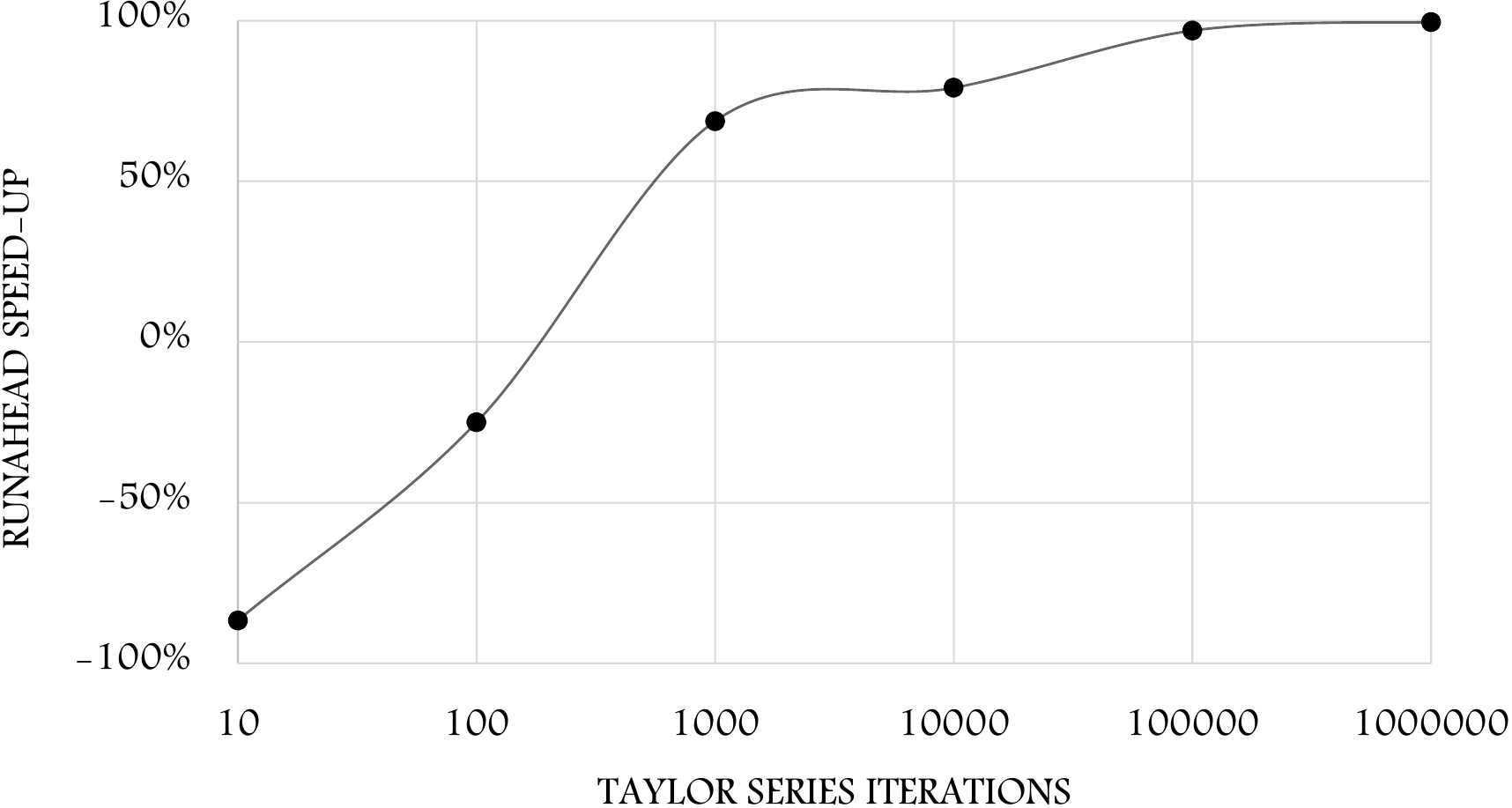}
\centering
\caption{Sensitivity of speed-up to the execution latency of input function in CPU platform.}
\label{fig6:cpu-speedup}
\end{figure}

Figure~\ref{fig6:cpu-speedup} shows the result of this study for CPU program. As the figure illustrates, when the execution latency of the function is low (below 500 iterations), the Runahead Computing not only does not improve the performance but also decreases it. This occurs because, in this situation, the overhead of creating and joining threads is more than the latency of computing the value of a point. But when the computation time of function goes beyond of a threshold, Runahead Computing improves the performance. In our experiment, Runahead Computing decreases the performance by 86\% when the latency of input function is small (10 iterations for Taylor series). By increasing the latency of function, the speed-up of Runahead Computing also increases. By setting the number of iterations of Taylor series to 10000, Runahead Computing improves the performance by 97\% and converges to the ideal speed-up value.\newline

\begin{figure}[t]
\includegraphics[width=0.4\textwidth]{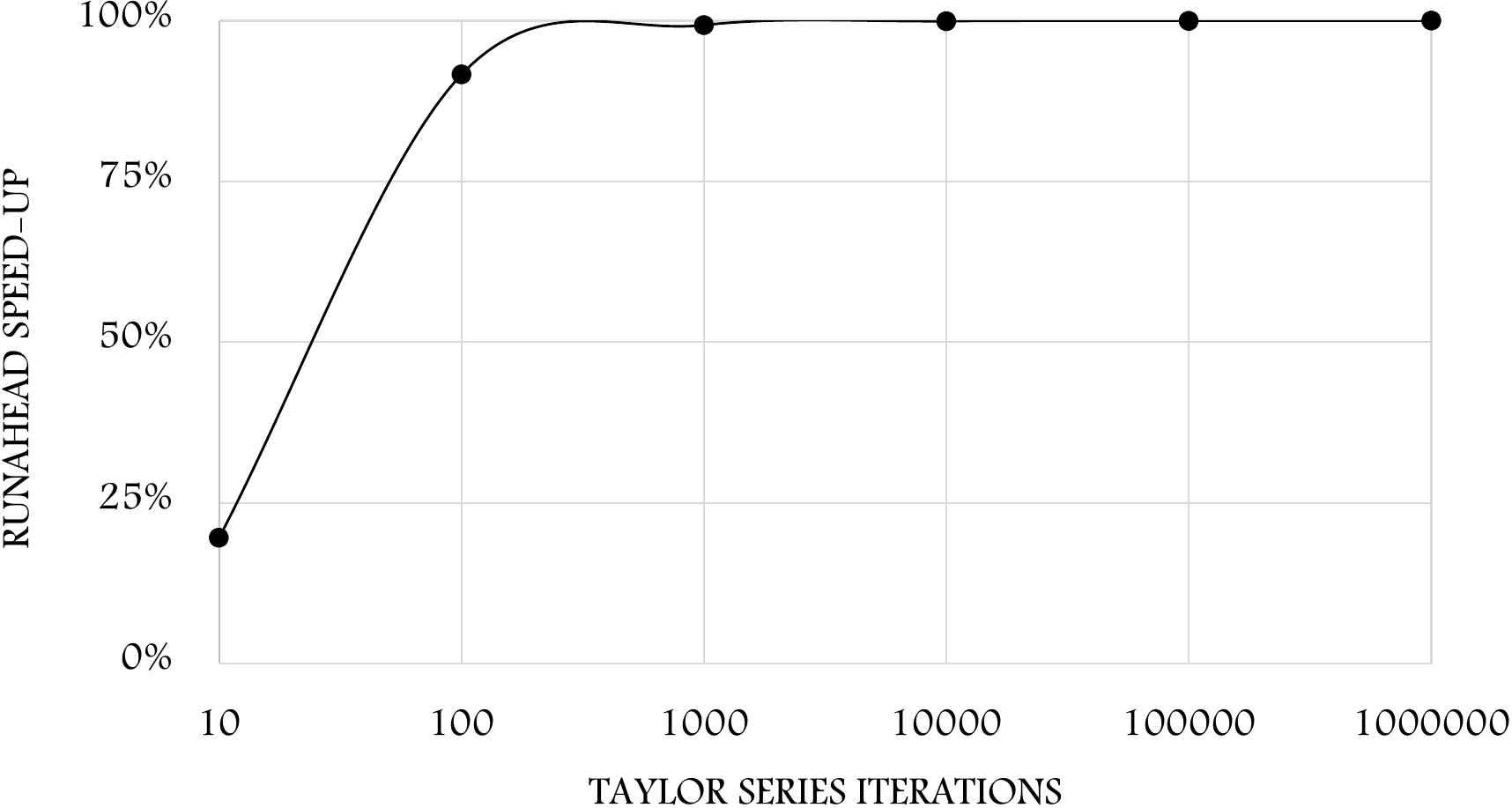}
\centering
\caption{Sensitivity of speed-up to the execution latency of input function in GPU platform.}
\label{fig7:gpu-speedup}
\end{figure}

Figure~\ref{fig7:gpu-speedup} presents the result of the same experiment on GPU platform. As shown, unlike CPU, Runahead Computing in the GPU never loses performance in our analysis. The reason is, the overhead of creating and joining the threads on GPU is very low in comparison with CPU~\cite{nvidia2008programming}. Even for a small number of iterations for Taylor series, Runahead Computing considerably increases the performance. For 10 iterations of Taylor series, Runahead Computing speeds up the execution latency by 19\%. For iterations beyond 500, Runahead Computing improves the performance by 99\% and reaches to its ideal speed-up value. \newline

\section{Conclusion}
In this paper, we proposed Runahead Computing, a technique for increasing the performance of single-threaded applications in multi-threaded architectures, by exploiting available idle threads. In the proposed approach, programmer codes the idle threads for working some steps ahead of the main thread. As a case study, we chose Bisection root-finding algorithm and accelerated it on a CMP and a GPU. While we examined our method on a particular algorithm, we believe that ideas in this paper can be applied to other similar algorithms (e.g., Binary Search).

\medskip
\bibliographystyle{unsrt}
\bibliography{myrefs}



\ifCLASSOPTIONcaptionsoff
  \newpage
\fi

\end{document}